\documentclass[aps,prl,preprint,tightenlines,superscriptaddress,showpacs,byrevtex]{revtex4}
\usepackage{graphicx} 
\usepackage{dcolumn}  
\usepackage{amsmath} 
\usepackage{color}
\usepackage{epsfig}
\usepackage{epstopdf}

\begin{document}

\preprint{\vbox{ \hbox{   }
                 \hbox{BELLE-CONF-0834}
}}

\title{ \quad\\[0.5cm]   \boldmath Measurements of time-dependent $CP$ Asymmetries in $B \to D^{*\mp} 
\pi^{\pm}$ decays using a partial reconstruction technique
}

\affiliation{Budker Institute of Nuclear Physics, Novosibirsk}
\affiliation{Chiba University, Chiba}
\affiliation{University of Cincinnati, Cincinnati, Ohio 45221}
\affiliation{Department of Physics, Fu Jen Catholic University, Taipei}
\affiliation{Justus-Liebig-Universit\"at Gie\ss{}en, Gie\ss{}en}
\affiliation{The Graduate University for Advanced Studies, Hayama}
\affiliation{Gyeongsang National University, Chinju}
\affiliation{Hanyang University, Seoul}
\affiliation{University of Hawaii, Honolulu, Hawaii 96822}
\affiliation{High Energy Accelerator Research Organization (KEK), Tsukuba}
\affiliation{Hiroshima Institute of Technology, Hiroshima}
\affiliation{University of Illinois at Urbana-Champaign, Urbana, Illinois 61801}
\affiliation{Institute of High Energy Physics, Chinese Academy of Sciences, Beijing}
\affiliation{Institute of High Energy Physics, Vienna}
\affiliation{Institute of High Energy Physics, Protvino}
\affiliation{Institute for Theoretical and Experimental Physics, Moscow}
\affiliation{J. Stefan Institute, Ljubljana}
\affiliation{Kanagawa University, Yokohama}
\affiliation{Korea University, Seoul}
\affiliation{Kyoto University, Kyoto}
\affiliation{Kyungpook National University, Taegu}
\affiliation{\'Ecole Polytechnique F\'ed\'erale de Lausanne (EPFL), Lausanne}
\affiliation{Faculty of Mathematics and Physics, University of Ljubljana, Ljubljana}
\affiliation{University of Maribor, Maribor}
\affiliation{University of Melbourne, School of Physics, Victoria 3010}
\affiliation{Nagoya University, Nagoya}
\affiliation{Nara Women's University, Nara}
\affiliation{National Central University, Chung-li}
\affiliation{National United University, Miao Li}
\affiliation{Department of Physics, National Taiwan University, Taipei}
\affiliation{H. Niewodniczanski Institute of Nuclear Physics, Krakow}
\affiliation{Nippon Dental University, Niigata}
\affiliation{Niigata University, Niigata}
\affiliation{University of Nova Gorica, Nova Gorica}
\affiliation{Osaka City University, Osaka}
\affiliation{Osaka University, Osaka}
\affiliation{Panjab University, Chandigarh}
\affiliation{Peking University, Beijing}
\affiliation{Princeton University, Princeton, New Jersey 08544}
\affiliation{RIKEN BNL Research Center, Upton, New York 11973}
\affiliation{Saga University, Saga}
\affiliation{University of Science and Technology of China, Hefei}
\affiliation{Seoul National University, Seoul}
\affiliation{Shinshu University, Nagano}
\affiliation{Sungkyunkwan University, Suwon}
\affiliation{University of Sydney, Sydney, New South Wales}
\affiliation{Tata Institute of Fundamental Research, Mumbai}
\affiliation{Toho University, Funabashi}
\affiliation{Tohoku Gakuin University, Tagajo}
\affiliation{Tohoku University, Sendai}
\affiliation{Department of Physics, University of Tokyo, Tokyo}
\affiliation{Tokyo Institute of Technology, Tokyo}
\affiliation{Tokyo Metropolitan University, Tokyo}
\affiliation{Tokyo University of Agriculture and Technology, Tokyo}
\affiliation{Toyama National College of Maritime Technology, Toyama}
\affiliation{Virginia Polytechnic Institute and State University, Blacksburg, Virginia 24061}
\affiliation{Yonsei University, Seoul}
  \author{I.~Adachi}\affiliation{High Energy Accelerator Research Organization (KEK), Tsukuba} 
  \author{H.~Aihara}\affiliation{Department of Physics, University of Tokyo, Tokyo} 
  \author{D.~Anipko}\affiliation{Budker Institute of Nuclear Physics, Novosibirsk} 
  \author{K.~Arinstein}\affiliation{Budker Institute of Nuclear Physics, Novosibirsk} 
  \author{T.~Aso}\affiliation{Toyama National College of Maritime Technology, Toyama} 
  \author{V.~Aulchenko}\affiliation{Budker Institute of Nuclear Physics, Novosibirsk} 
  \author{T.~Aushev}\affiliation{\'Ecole Polytechnique F\'ed\'erale de Lausanne (EPFL), Lausanne}\affiliation{Institute for Theoretical and Experimental Physics, Moscow} 
  \author{T.~Aziz}\affiliation{Tata Institute of Fundamental Research, Mumbai} 
  \author{S.~Bahinipati}\affiliation{University of Cincinnati, Cincinnati, Ohio 45221} 
  \author{A.~M.~Bakich}\affiliation{University of Sydney, Sydney, New South Wales} 
% ...
\collaboration{The Belle Collaboration}

\begin{abstract}
We report preliminary results on time-dependent $CP$ asymmetries in $B \to D^{*\mp}\pi^{\pm}$ decays. The $CP$ asymmetry in these decays is proportional to $2R_{D^{*} \pi} \sin (2\phi_1 + \phi_3 \pm \delta_{D^{*} \pi})$, where $R_{D^{*} \pi}$ is the ratio of the magnitudes of the doubly-Cabibbo-suppressed and Cabibbo-favoured amplitudes, $\delta_{D^{*} \pi}$ is the strong phase difference between them, and $\phi_1$ and $\phi_3$ are two angles of the CKM Unitarity Triangle. This study is based on a large data sample that contains 657 million $B\overline{B}$ pairs collected with the Belle detector at the KEKB asymmetric-energy $e^+ e^-$ collider at the $\Upsilon(4S)$ resonance. We use a partial reconstruction technique, wherein signal $B \to D^{*\mp}\pi^{\pm}$ events are identified using information only from the $\pi^{\pm}$ from the $B$ decay and the charged slow pion from the subsequent decay of the $D^{*-}$. We obtain 
  $S^+ (D^* \pi) = +0.057 \pm 0.019(\mathrm{stat}) \pm 0.012(\mathrm{sys})$ and
  $S^- (D^* \pi) = +0.038 \pm 0.020(\mathrm{stat}) \pm 0.010(\mathrm{sys})$. 
  \end{abstract}

\maketitle

\tighten

{\renewcommand{\thefootnote}{\fnsymbol{footnote}}}
\setcounter{footnote}{0}

\section{Introduction}

\begin{figure}[!htb]
 \includegraphics[width=16.0cm,clip]{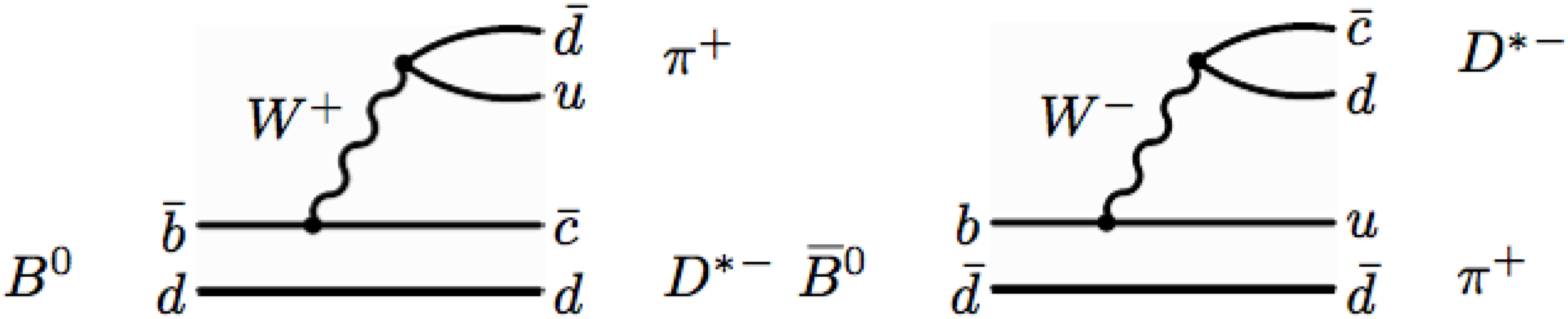}   
    \caption{
      Diagrams for 
       $B^0 \to D^{*-}\pi^+$ (left) and 
      $\overline{B}{}^0 \to D^{*-}\pi^+$ (right).
      Those for $\overline{B}{}^0 \to D^{*+}\pi^-$ and 
      $B^0 \to D^{*+}\pi^-$
      can be obtained by charge conjugation.}
      \label{fig:feynman}
\end{figure}

In the Standard Model (SM), quark flavour mixing occurs via the Cabibbo-Kobayashi-Maskawa (CKM) matrix~\cite{KM}. $CP$ violation in the SM occurs due to the presence of a complex phase in the CKM matrix. Precision measurements of the parameters of CKM matrix are of utmost importance to constrain the SM and measure the amount of $CP$ violation. The study of the time-dependent decay rates of $B^0 (\overline{B}{}^0) \to D^{*\mp}\pi^{\pm}$ provide a theoretically clean method for extracting $\sin(2\phi_1+\phi_3)$~\cite{dunietz}, where $\phi_1$ and $\phi_3$ are angles of the CKM Unitarity Triangle. As shown in Fig.~\ref{fig:feynman}, this decay can be mediated by both Cabibbo-favoured (CFD) and doubly-Cabibbo-suppressed (DCSD) processes, whose  
amplitudes are proportional to $V_{cb}^*V_{ud}$ and $V_{ub}^*V_{cd}$ respectively, which have a relative weak phase $\phi_3$. 

The time-dependent decay rates are given by~\cite{fleischer}
\begin{eqnarray}
P(B^{0} &\to& D^{(*)\pm} \pi^\mp) = \frac{1}{8\tau_{B^0}} 
                    e^{-|\Delta t|/\tau_{B^0}} 
   \times \left[
      1 \mp C \cos (\Delta m \Delta t) - S^\pm \sin (\Delta m \Delta t) 
    \right],  \nonumber \\
P(\overline{B}{}^0 &\to& D^{(*)\pm} \pi^\mp) =  \frac{1}{8\tau_{B^0}}  
                  e^{-|\Delta t|/\tau_{B^0}} 
  \times    \left[
      1 \pm C \cos (\Delta m \Delta t) + S^\pm \sin (\Delta m \Delta t) 
    \right]. 
        \label{eq:evol}  
  \end{eqnarray}
Here $\Delta t$ is the difference between the time of the decay and the 
time that the flavour of the $B$ meson is tagged,  
$\tau_{B^0}$ is the average neutral $B$ meson lifetime, 
$\Delta m$ is the $B^0$-$\overline{B}{}^0$ mixing parameter, and 
$C = \left( 1 - R^2 \right) / \left( 1 + R^2 \right)$, 
where $R$ is the ratio of the magnitudes between the DCSD and CFD 
(we assume the magnitudes of both the CFD and  DCSD amplitudes are the
same for $B^0$ and $\overline{B}{}^0$ decays). 
The $CP$ violation parameters are given by
\begin{equation}
S^{\pm} = \frac{2 (-1)^L R \sin(2\phi_1+\phi_3 \pm \delta)}
               { \left( 1 + R^2 \right)},
\label{eq:spm}
\end{equation}
where $L$ is the orbital angular momentum of
the final state (1 for $D^* \pi$), and $\delta$ is 
the strong phase difference between CFD and DCSD. Since the predicted value of $R$ is small, $\sim$ 
$0.02$~\cite{csr}, we neglect terms of ${\cal O}\left( R^2 \right)$ 
(and hence take $C = 1$).  The strong phase $\delta$ for $D^*\pi$ is predicted to be small~\cite{fleischer,wolfenstein}. Since $R$ is expected to be suppressed, the amount of $CP$ violation in $D^*\pi$ decays, which is proportional to $R$, is expected to be small and a large data sample is needed in order to
obtain sufficient sensitivity. We employ a partial 
reconstruction technique~\cite{zheng} for the $D^* \pi$ analysis, wherein the signal is distinguished from background on the basis of 
kinematics of the `fast' pion ($\pi_f$) from the decay $B \to D^* \pi_f$,
and the `slow' pion (`$\pi_s$')  from the subsequent decay of  $D^* \to D \pi_s$;
the $D$ meson is not reconstructed at all. 

     Previous analyses have been reported by Belle~\cite{belle_partial, fronga} as well as by Babar~\cite{babar_partial}. This study uses a data sample of $605\,\mathrm{fb}^{-1}$ containing 657 million $B \overline{B}$ events, which is about two times the size of the dataset used in the previous Belle analysis ~\cite{fronga} and supersedes the previous Belle result.

\section{Belle Detector}
The data were collected with the Belle
detector~\cite{Belle} at the KEKB asymmetric energy electron-positron ($e^-e^+$)
collider~\cite{KEKB} operating at the $\Upsilon$(4S) resonance. 
The Belle detector is a large-solid-angle magnetic 
spectrometer that consists of a silicon vertex detector (SVD), 
a 50-layer central drift chamber (CDC), an array of aerogel 
threshold Cherenkov counters (ACC), a barrel-like arrangement 
of time-of-flight scintillation counters (TOF), and an 
electromagnetic calorimeter (ECL) comprised of CsI(Tl) 
crystals located inside a superconducting solenoidal coil 
that provides a 1.5 T magnetic field. An iron flux-return 
located outside of the coil is instrumented to detect $K_L^0$ 
mesons and to identify muons (KLM). 
A sample containing 
152 million $B \overline{B}$ pairs was collected with 
a 2.0~cm radius beampipe and a 3-layer silicon vertex detector 
(SVD1), 
while a sample of 505 million $B \overline{B}$ pairs was 
collected with a 1.5~cm radius beampipe, a 4-layer silicon vertex
detector (SVD2), and a small-cell inner drift chamber~\cite{svd2}. 

\section{Analysis procedure}
\subsection{Partial Reconstruction of $B \to D^{*\mp}\pi^{\pm}$ decays}

To reconstruct the $CP$-side tags, we use:  $\overline{B} \rightarrow D^{*\pm}\pi^{\mp}$;  $D^{*\pm} \to D^{0} \pi^{\pm}$. Candidate events are selected by requiring the presence of oppositely charged `\textit{f}' and `\textit{s}' candidates. We estimate the $D^{*}$ frame using energy-momentum conservation:
\begin{eqnarray}
 E_{D^{*}} = E_{\overline{B}{}^0} - E_{f}, \nonumber \\
 \vec{p}_{f} + \vec{p}_{D^{*}} = \vec{p}_{B}.
\label{eqnpr}
 \end{eqnarray} 
 Here, $E$ and $p$ stand for energy and momentum respectively. $E_{\overline{B}{}^0}$ is half the total centre-of-mass energy ($E_{CM}$) of the incoming $e^+e^-$ beams and $p_B = \sqrt {{E^2_{CM}}/4 - {m_{B^{0}}^2}}$, where $m_{B^0}$ is the nominal $B^0$ mass~\cite{PDG}.
Using $E_{D^{*}}$, the momentum of $D^{*}$ can be obtained in the $e^+e^-$ centre-of-mass (cms) as: 
$p_{D^{*}}$ = $\sqrt{E_{D^{*}}^{2} - m_{D^{*}}^{2}}$. We construct a partially reconstructed $D^{*+}$ frame, using $p_{D^{*}}$ and  $E_{D^{*}}$ and taking the direction of $\vec{p}_{D^{*}}$ opposite to $\vec{p}_{f}$.

We define a variable, $p_{\delta}$, which is strongly correlated with the fast pion momentum, ${p}_{f}$.
 $p_{\delta}$ is defined as: $||{p}_{f}| - |{p}_{D^{*}}||$ and from Eq.~(\ref{eqnpr}), it follows: 
$|p_{\delta}| \le |\vec{p}_{B}|$ ($\approx$ 0.3 GeV$/c$). We boost the charged slow pion into the partially reconstructed $D^{*}$ frame. In the true $D^{*}$ frame, 
the slow pion is mono-energetic. However, in the partially reconstructed $D^{*+}$ frame, 
the slow pion momentum will have a limited spread.  We study the parallel and the transverse 
components of the momentum of the slow pion, ${\pi_{s}}$
 along the direction opposite to $f$, which are denoted as ${p}_{\parallel}$ and 
${p}_{\perp}$, respectively. In the true $D^{*}$ frame, the ${p}_{\parallel}$ variable has a distribution proportional to $\cos \theta^2$ for signal 
events,
as the $B$ decay is a pseudoscalar to pseudo-scalar vector transition. 

\subsection{$CP$ side selection}
 Fast pion ($f$) candidates are required to have a radial (longitudinal) impact parameter $dr <
0.1\,\mathrm{cm}$ ($|dz| < 2.0\,\mathrm{cm}$) and
to have associated hits in the SVD. We reject leptons and kaons based on information from the CDC, TOF and ACC from the fast pion candidate list.
A requirement is made on the fast pion cms momentum, $1.93 \, {\rm GeV}/c < p_{f} < 2.50 \, {\rm GeV}/c$. Slow pion ($s$) candidates are required to have cms momentum in the range
$0.05 \, {\rm GeV}/c < p_{s} < 0.30 \, {\rm GeV}/c$. Since slow pions are not used for vertexing, no particle  identification requirement is applied, we impose only a loose requirement that they originate from the IP. We select candidates that satisfy
$ +0.00 \, {\rm GeV}/c < p_\perp    < +0.06 \, {\rm GeV}/c$,  $-0.10 \, {\rm GeV}/c < p_\parallel < 0.07 \, {\rm GeV}/c$ and $-0.60 \, {\rm GeV}/c < p_\delta  < 0.50 \, {\rm GeV}/c$ .

\subsection{Vertexing and Flavour Tagging}
The determination of the flavour of the $B$ meson opposite to the $CP$-side $B$ is essential for the $\Delta t$ measurement. In order to tag the flavour of the associated $B$ meson,
we require the presence of a high-momentum lepton ($l$) in the event. This helps reduce 
background from continuum 
$e^+e^- \to q\overline{q} \ (q = u,d,s,c)$ processes.
Tagging lepton candidates are required to be positively identified
either as electrons, on the basis of information from the CDC, ECL and ACC, 
or as muons, on the basis of information from the CDC and the KLM.
They are required to have momenta in the range
$1.1 \ {\rm GeV}/c < p_{l} < 2.3 \ {\rm GeV}/c$,
and to have an angle with the fast pion candidate that satisfies
$-0.75 < \cos \delta_{\pi_f l}$ in the cms.
The lower bound on the momentum and the requirement on the angle 
also reduce, to a negligible level (0.7$\%$), the contribution of leptons produced from semi-leptonic
decays of the unreconstructed $D$ mesons in the $B^0 \to D^{*\mp}\pi^\pm$ decay chain.

Identical vertexing requirements to those for fast pion candidates 
are made in order to obtain an accurate $z_{\rm tag}$ position.
To further suppress the small remaining continuum background,
we impose a loose requirement on the ratio of 
the second to zeroth Fox-Wolfram~\cite{fw} moments, $R_2 < 0.6$.

 At the KEKB asymmetric-energy $e^{+}e^{-}$ (3.5 GeV on 8 GeV) collider, operating at the $\Upsilon(4S)$ resonance 
($\sqrt s$ = 10.58 GeV), the $\Upsilon(4S)$ is produced with a Lorentz boost of $\beta\gamma$ = 0.425, almost
along the electron beamline ($z$) at KEKB. In the cms, $B^{0}$ and $\overline{B}{}^0$ mesons are
approximately at rest. Hence the proper time-difference ($\Delta t$) between the $z_{CP}$ and $z_{\rm tag}$ vertices is obtained from fast pion on the $CP$-side and the tagging lepton. The variable $\Delta t$ is defined as:
\begin{eqnarray}
 \Delta t \approx (z_{CP} - z_{\rm tag}) /\beta\gamma c.
\label{eqn2}
\end{eqnarray} 
The $CP$-side ($z_{CP}$) vertex is obtained from the fast pion on the $CP$-side and the run-dependent interaction point profile (IP). The tag-side ($z_{\rm tag}$) vertex is obtained from tagging lepton and the run-dependent IP.

\subsection{Yield Fit}
We use the three
kinematic variables, $p_{\delta}$,
${p}_{\parallel}$ and ${p}_{\perp}$ to distinguish between 
signal and background on the $CP$ side. Background events are separated into three categories:
$D^{*\mp}\rho^{\pm}$, which is kinematically similar to the signal; 
correlated background, in which the slow pion originates from the decay of 
a $D^*$ that originates from the decay of the same $B$ as the fast pion candidate ({\it e.g.}, $D^{**}\pi$);
and uncorrelated background, which includes everything else
({\it e.g.}, continuum processes, $D\pi$). The kinematic distributions of the signal and background categories 
are determined from a large MC sample,
corresponding to three times the integrated luminosity of our data sample.

  We select candidates that satisfy
$-0.10 \, {\rm GeV}/c < p_\parallel < +0.07 \, {\rm GeV}/c$ and $-0.60 \, {\rm GeV}/c < p_\delta < +0.50 \, {\rm GeV}/c$.
In the cases where more than one candidate satisfies these criteria,
we select the one with the largest value of $\delta_{\pi_f \pi_s}$, where $\delta_{\pi_f \pi_s}$ is the angle between the fast pion direction and the slow pion direction in the cms frame. The signal region is defined as: $-0.40 \, {\rm GeV}/c < p_\delta <  +0.40 \, {\rm GeV}/c$ and two regions in ${p}_{\parallel}$: 
$-0.05 \, {\rm GeV}/c < p_\parallel < -0.01 \, {\rm GeV}/c$ and $+0.01 \, {\rm GeV}/c < p_\parallel  <  +0.04 \, {\rm GeV}/c$.

\begin{figure}[!htb]
\includegraphics[width=17.0cm,clip]{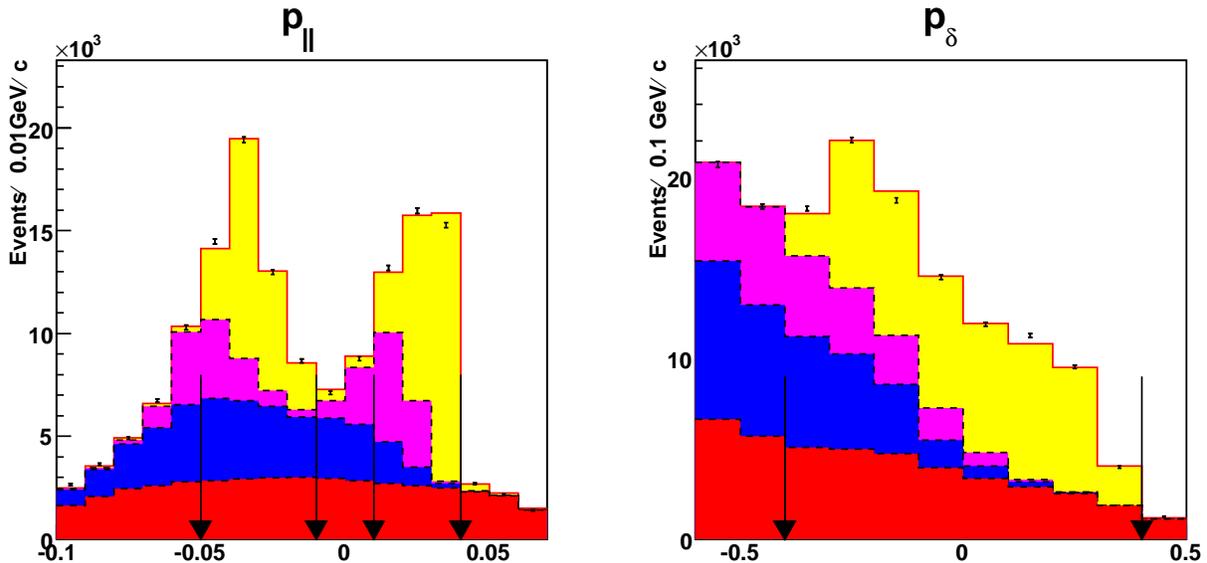}   
  \caption{
    \label{fig:kin_fit}
    Results of the yield fits to $D^*\pi$ candidates projected onto the $p_{\parallel}$ (left) and $p_{\delta}$ (right) axes in the signal region of the kinematic variables. The contributions are: $D^* \pi$ (yellow), $D^* \rho$ (magenta), correlated background (blue) and uncorrelated background (red).
  }
\end{figure}

Event-by-event signal and background fractions are determined from 
binned maximum likelihood fits to the two-dimensional 
kinematic  distributions ($p_\delta$ and $p_\parallel$). 
The results of these fits, projected onto each of the two variables,
are shown in Fig.~\ref{fig:kin_fit},
and summarized in Table~\ref{tab:kin_fit}. 
We obtain a signal purity of 
$59.0\pm0.4\%$, where purity is defined as ratio of the signal and total yield.

\begin{table}[htb]
   \caption{
     \label{tab:kin_fit}
     Summary of the yields in the signal region
        }
   \begin{center}
     \begin{tabular}{lc}
\hline 
              &  Candidates  \\
\hline 
$D^* \pi$    & $ 50196\pm286 $  \\
$D^* \rho$   & $  10232\pm 150 $ \\
Correlated background   & $  10425\pm 135 $   \\
Uncorrelated background    & $  14193\pm 128 $  \\
\hline 
     \end{tabular}
   \end{center}
\end{table}

\subsection{\boldmath Fit Procedure to obtain $CP$ violation parameters}
We perform a simultaneous unbinned fit to the 
same-flavour (SF) events, in which the fast pion and the tagging lepton
have the same charge, and opposite-flavour (OF) events, in which the
fast pion ($f$) and the tagging lepton ($l$) have the opposite charge, to measure the $CP$ violation parameters in the $D^*\pi$ sample. We minimize the quantity $-2\ln {\cal L} = -2 \sum_i \ln {\cal L}_i$,
where 
\begin{equation}
  \label{eq:likelihood}
  {\cal L}_i = 
  f_{D^*\pi} P_{D^*\pi} + f_{D^*\rho} P_{D^*\rho} +
  f_{\rm unco} P_{\rm unco} + f_{\rm corr} P_{\rm corr}.
\end{equation}
Here, $f$ stands for the event-by-event signal and background fractions
and are obtained from the fits to the kinematic variables and $P$ stands for the probability density functions (PDF) for signal and backgrounds, which contain a physics PDF and experimental effects. For $D^*\pi$ and $D^*\rho$, the PDF is given by Eq.~(\ref{eq:evol}),
where for $D^*\rho$ the terms $S^\pm$ are effective parameters
averaged over the helicity states~\cite{dstarrho} and are constrained to be zero.
The PDF for correlated background contains
a term for neutral $B$ decays 
(given by Eq.~(\ref{eq:evol}) with $S^\pm = 0$),
and a term for charged $B$ decays
(for which the PDF is 
$\frac{1}{2\tau_{B^+}} e^{-\left| \Delta t \right| / \tau_{B^+}}$,
where $\tau_{B^+}$ is the lifetime of the charged $B$ meson). The PDF for uncorrelated background also contains
neutral and charged $B$ components, with the remainder from continuum 
$e^+e^- \to q\overline{q} \ (q = u,d,s,c)$ processes.
The continuum PDF is modelled with two components: 
one with negligible lifetime,  and the other with a finite lifetime.

The parameters in $P_{\rm unco}$ and $P_{\rm corr}$ are obtained from separate simultaneous fits to OF and SF candidates in the respective sideband regions. In these fits, the $CP$ violation parameters are fixed to 0 since there is no $CP$ in background. The fit is further simplified by fixing the biases in $\Delta z$ to zero (discussed later in detail).  Monte-Carlo simulation studies demonstrate that floating or fixing these biases to 0 does not affect the background parameters.
  
    To measure the uncorrelated background shape,
we use events in a sideband region, $-0.10 \, {\rm GeV}/c < p_{\parallel} < 0.07 \, {\rm GeV}/c$, $0.01 \, {\rm GeV}/c < p_{\parallel} < 0.04 \, {\rm GeV}/c$, $-0.60 \, {\rm GeV}/c < p_{\delta} < 0.50 \, {\rm GeV}/c$ and $0.08 \, {\rm GeV}/c <p_{\perp} < 0.10 \, {\rm GeV}/c$, which is populated mostly by uncorrelated background. To determine the correlated background parameters, we use events 
in a sideband region,
$-0.10 \, {\rm GeV}/c < p_{\parallel} < 0.07 \, {\rm GeV}/c$,
$-0.60 \, {\rm GeV}/c < p_{\delta}< 0.00 \, {\rm GeV}/c$ and
$0.00 \, {\rm GeV}/c < p_{\perp} < 0.05 \, {\rm GeV}/c$. This sideband region is dominated by correlated and 
uncorrelated backgrounds and has very small amount of $D^*\pi$ signal and $D^*\rho$ background. The uncorrelated background parameters are fixed to the values
obtained in the previous fit.

    The PDF for the signal and background in Eq.~(\ref{eq:likelihood})
must be convolved with corresponding $\Delta z$ resolution functions related to 
kinematic smearing (${\mathcal{R}_{\rm k}}$), detector resolution (${\mathcal{R}_{\rm det}}$), and asymmetry in $\Delta z$ due to non-primary tracks (${\mathcal{R}_{\rm np}}$). The resolution function related to kinematic smearing is due to the fact that we use the approximation of Eq.~(\ref{eqn2}). The detector resolution function parameters are obtained using $J/\psi \to \mu^+\mu^-$ candidates. Since both the fast pion and 
the tagging lepton originate directly from $B$ meson decays for correctly tagged signal events, we do not include any additional smearing due to non-primary 
tracks in such events. However, for incorrectly tagged events, almost exclusively
originating from secondary leptons or pions, the PDF is 
convolved with an additional resolution component whose parameters
are determined from MC simulations. The detector resolution and smearing due to asymmetry in $\Delta z$ due to non-primary tracks are described in detail elsewhere~\cite{fronga}.

Mistagging is taken into account using
\begin{eqnarray}
    \label{eq:exp_pdf}
   P( l_\mathrm{tag}^{\mp}, \pi_f^\pm)  = 
   ( 1 - w_{\mp} ) P(B^{0} /\overline{B}{}^0 \to D^{*\mp} \pi^\pm)       
  +  w_{\pm} P(\overline{B}{}^0/ B^{0} \to D^{*\mp} \pi^\pm) 
 \end{eqnarray}
where $\pi_f$ is fast pion in $CP$-side, $l$ is tag-side lepton, $w^{+}$ and $w^{-}$ are the wrong-tag fractions, defined as the probabilities to incorrectly measure the flavour of tagging $B^{0}$ and $\overline{B}{}^0$ mesons respectively and are determined from the data as free parameters in the fit for $S^\pm$.

The time difference $\Delta t$ is related to the measured quantity $\Delta z$
as described in Eq.~(\ref{eqn2}), with an additional term due to 
possible offsets in the mean value of $\Delta z$,
\begin{equation}
  \label{eq:dt_offset}
  \Delta t \longrightarrow \Delta t + \epsilon_{\Delta t} \simeq \left( \Delta z + \epsilon_{\Delta z} \right) / \beta\gamma c.
\end{equation}
It is essential to allow non-zero values of $\epsilon_{\Delta t}$ since a 
small bias can mimic the effect of $CP$ violation:
\begin{equation}
      \cos (\Delta m \Delta t) 
 \to
      \cos (\Delta m \Delta t) - 
\Delta m \epsilon_{\Delta t} \sin (\Delta m \Delta t)
\end{equation}
A bias as small as $\epsilon_{\Delta z} \sim 1 \ \mu{\rm m}$ can lead to 
sine-like terms as large as $0.01$, 
comparable to the expected size of the $CP$ violation effect.
 Because both vertex positions are obtained from single tracks,  
the partial reconstruction analysis is more susceptible than other Belle $CP$ analyses to 
such biases. We allow separate offsets for each combination of $h$ and $l$ charges. In order to correct for a known bias due to the relative misalignment of the SVD 
and CDC in SVD1 data, a small correction is applied to each measured vertex position. This correction is dependent on the track charge, 
momentum and polar angle, measured in the laboratory frame and is obtained by comparing the vertex positions calculated with the alignment constants used in the data,
to those obtained with an improved set of alignment 
constants~\cite{dzb},  which removes the observed bias. 
Since the alignment in SVD2 data was found to be comparable to that of
the corrected SVD1 data, no additional correction was applied to SVD2 data.

\subsection{Fit Result}
In order to test our fit procedure,
we first constrain $S^+$ and $S^-$ to be zero and perform a fit in
which $\tau_{B^0}$ and $\Delta m$ 
(as well as two wrong tag fractions and eight offsets) are free parameters.
We obtain $\tau_{B^0} = 1.538 \pm 0.008 \ {\rm ps}$
and $\Delta m = 0.482 \pm 0.004 \ {\rm ps}^{-1}$,
where the errors are statistical only.
These values are compatible with their world average values~\cite{PDG}.
Reasonable agreement with the input values is also obtained in MC.
Furthermore, fits to the MC with $S^\pm$ floated give results 
consistent with zero, as expected.

\begin{figure}[!htb]
  \begin{center}
    \includegraphics[width=14.0cm,clip]{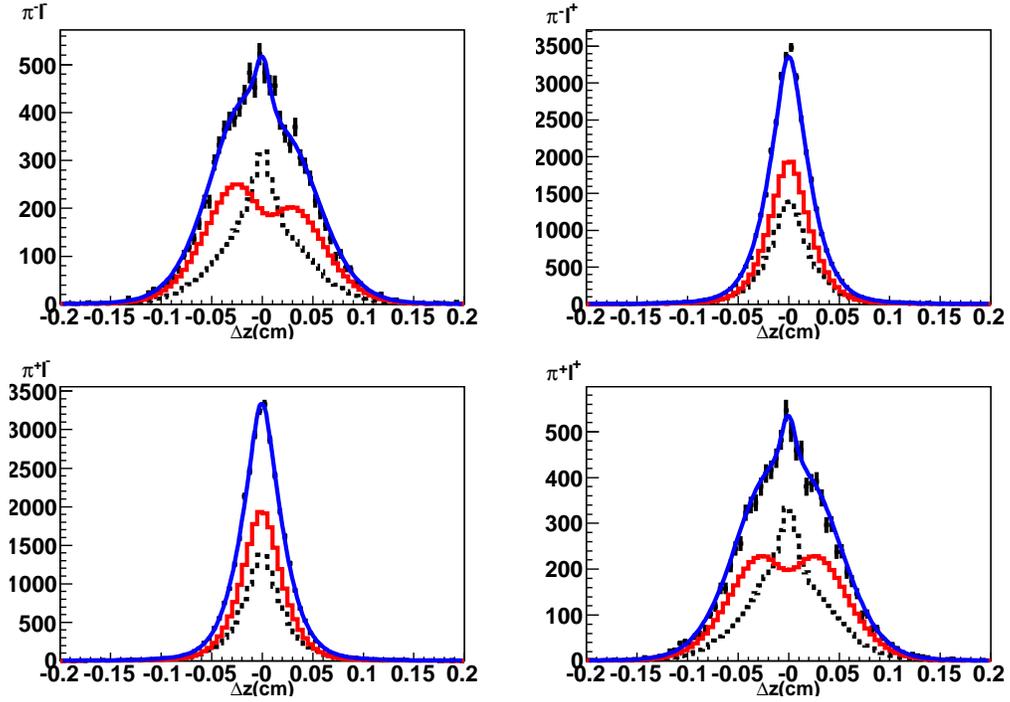}

  \end{center}
  \caption{
    \label{fig:myfit}
    $\Delta z$ distributions for 4 flavour-charge combinations: $\pi^{-}l^{-}$ (top left) , $\pi^{-}l^{+}$ (top right), $\pi^{+}l^{-}$ (bottom left), $\pi^{+}l^{+}$ (bottom right). The fit result is superimposed on the data (blue line).
    The signal and background components are shown as the red and dotted black curves, respectively.
   }
\end{figure}

\begin{figure}[!htb]
  \begin{center}
    \includegraphics[width=14.cm,clip]{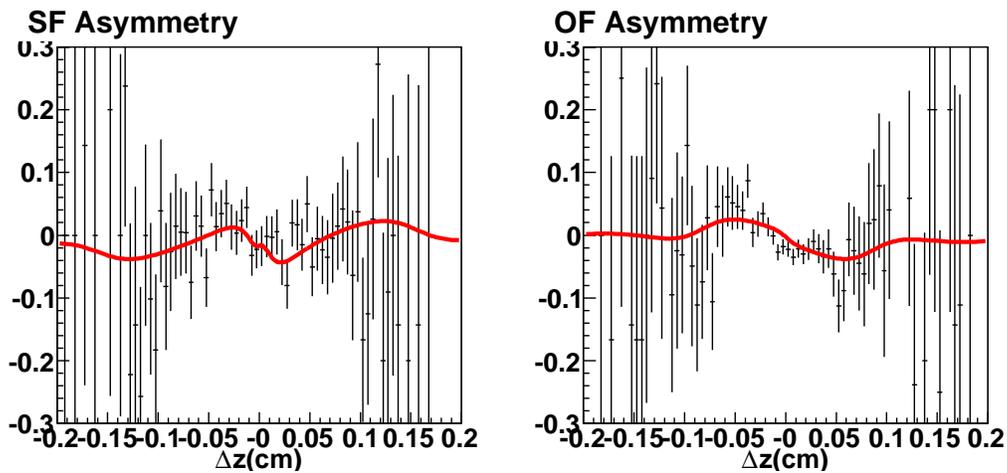}
  \end{center}
  \caption{
    \label{fig:myfit_asp}
    Results of the fit to obtain $S^+$ and $S^-$,
    shown as asymmetries in the same flavour events (left)  and opposite flavour events (right).
    The fit result (red curve) is superimposed on the data.
  }
\end{figure}

To extract the $CP$ violation parameters we fix $\tau_{B^0}$ and 
$\Delta m$ at their world average values,
and fit with $S^+$, $S^-$, two wrong tag fractions, and eight offsets
as free parameters.
We obtain
$S^+ =  0.057\pm 0.019$ and $S^- =  0.038\pm 0.020$
where the errors are statistical only.
The wrong tag fractions are $w_- = (6.8 \pm 0.3)\%$ and 
$w_+ = (6.6 \pm 0.3)\%$. 
All floating offsets are consistent with zero except  for one of the OF combinations ($h = \pi^{-}$, $l = \ell^{+}$) in the SVD1 sample.
The results are shown in Fig.~\ref{fig:myfit}.
To further illustrate the $CP$ violation effect,
we define asymmetries in the 
same flavour events (${\cal A}^{\mathrm{SF}}$) 
and in the opposite flavour events (${\cal A}^{\mathrm{OF}}$), as 
\begin{eqnarray}
  {\cal A}^{\mathrm{SF}} & = &
  \frac{ N_{\pi^-l^-}(\Delta z) - N_{\pi^+l^+}(\Delta z) } 
       { N_{\pi^-l^-}(\Delta z) + N_{\pi^+l^+}(\Delta z) }, \nonumber \\
  {\cal A}^{\mathrm{OF}} & = &
  \frac{ N_{\pi^+l^-}(\Delta z) - N_{\pi^-l^+}(\Delta z) } 
       { N_{\pi^+l^-}(\Delta z) + N_{\pi^-l^+}(\Delta z) }, 
\end{eqnarray}
where the $N$ values indicate the number of events 
for each combination of $h$ and $l$ charge.
These are shown in Fig.~\ref{fig:myfit_asp}.

\subsection{Systematic Error}
This analysis is very sensitive to the vertexing bias.  Hence, we have use $\Delta z$ offsets in the fits to take care of this bias. In order to estimate the error due to these offsets, we use the difference of the mean of $S^{\pm}$ obtained using an ensemble of  300 generated $D^{*}\pi$ signal samples with $CP$ ($S^{\pm} = -0.04$)  and the generated $CP$ value. 

Other sources of systematic error are the parameters of resolution functions,  ${\mathcal{R}_{\rm k}}$, ${\mathcal{R}_{\rm det}}$ and ${\mathcal{R}_{\rm np}}$, the parameters of uncorrelated and correlated background and physics parameters, $\Delta m$,  $\tau_{B^0}$, $\tau_{B^{+}}$, $S^{\pm}_{D^{*}\rho}$ and $S^{\pm}_{\rm corr}$ that are fixed in the $CP$ fit, where $S^{\pm}_{\rm corr}$ are the $CP$ violation parameters for the correlated background component ($S^{\pm}_{\rm corr} = \pm 0.05$ in the $CP$ fit). Additional systematic errors can result from varying the number of bins for the kinematic variables, $p_{\delta}$ and ${p}_{\parallel}$ in the yield fit.

We use a triple Gaussian as the detector resolution ($R_{\rm det}$) function model. We consider the systematic uncertainty due to lack of knowledge of the exact functional form of the resolution model. Hence, we change the resolution models and obtain shifts as large as $0.008$ for $S^{+}$. This is conservatively assigned as the systematic error due to lack of knowledge of the resolution model.

We also performed a linearity test to check for possible fit bias by generating a number of large samples of signal MC simulations with different input values of $S^+$ and $S^-$. All results are
consistent with the input values, without evidence of any bias. In addition, we checked the pull for $S^+$ and $S^-$ using
two types of ensembles, one set generated with no $CP$ ($S^{\pm} = 0$) and the other with $CP$ ($S^{\pm} = -0.04$) and obtained mean ($m$) and rms ($\sigma$) of the pull distributions fitted to a single Gaussian. For the no $CP$ case, $m_{S^{+}} = +0.10 \pm 0.06$, $\sigma_{S^{+}} = +0.94 \pm 0.05$; $m_{S^{-}} = -0.10 \pm 0.06$, $\sigma_{S^{-}} = +0.96 \pm 0.05$ and for the case with $CP$, $m_{S^{+}} = +0.14 \pm 0.07$, $\sigma_{S^{+}} = +1.10 \pm 0.06$; $m_{S^{-}} = -0.29 \pm 0.06$, $\sigma_{S^{-}} = +0.98 \pm 0.05$. Both cases yield $m$ and $\sigma$ for the pull distributions close to 0 and 1, respectively. This shows that our fit routine does not have any significant bias.

The systematic errors are summarized in Table~\ref{tab:systematics}.
The total systematic error is obtained by adding the 
above terms in quadrature.
\begin{table}[htb]
  \begin{center}
 \begin{tabular}{lccc} \hline  
  Systematic error source& $ S^{+}$ &$S^{-}$ \\\hline 
  $\Delta z$ offset &$0.002$&$0.003$\\
   ${\mathcal{R}_{k}}$ parameters  &$0.002$&$0.003$\\
  ${\mathcal{R}_{det}}$ parameters&$0.002$&$0.002$\\
 ${\mathcal{R}_{np}}$ parameters &$0.008$&$0.007$\\
 Background parameters &$0.002$&$0.003$\\
 Physics parameters &$0.004$&$0.004$\\
 Yield fit &$0.003$&$0.003$\\
 Resolution model &$0.006$&$0.002$\\
$\Delta z$ floated in background PDF&$0.000$&$0.000$\\\hline
Total systematic error &$0.012$&$0.010$\\
    \hline 
    \end{tabular}
    \caption{Summary of possible sources of systematic error }
\label{tab:systematics}
  \end{center}
\end{table}

The results using the partial reconstruction method are
\begin{eqnarray}
 S^+ & = &  +0.057\pm 0.019 \pm 0.012 , \nonumber \\
 S^- & = & +0.038 \pm 0.020 \pm 0.010 , 
\end{eqnarray}
where the first error is statistical and the second error is systematic.

\section{Summary}
 
We have measured $CP$ violation parameters that depend on $\phi_3$ using the 
time-dependent decay rates of the decay $B^0 \to D^{*\mp} \pi^\pm$ using a data sample containing 657 million $B \overline{B}$ events. We obtain the $CP$ violation parameters expressed in terms of $S^+$ and $S^-$, which are
related to the CKM angles $\phi_1$ and $\phi_3$, the ratio of suppressed to 
favoured amplitudes, and the strong phase difference between them,  as 
$S^{\pm} = -R_{D^*\pi} \sin(2\phi_1+\phi_3 \pm \delta_{D^*\pi})/
                \left( 1 + R_{D^*\pi}^2 \right)$ for $D^* \pi$ as
                
\begin{eqnarray}
 S^+ & = &  +0.057\pm 0.019 \pm 0.012 ,  \nonumber \\
 S^- & = & +0.038 \pm 0.020 \pm 0.010 , 
\end{eqnarray}
where the first errors are statistical and the second errors are systematic.

\section*{Acknowledgements}
We thank the KEKB group for the excellent operation of the
accelerator, the KEK cryogenics group for the efficient
operation of the solenoid, and the KEK computer group and
the National Institute of Informatics for valuable computing
and SINET3 network support. We acknowledge support from
the Ministry of Education, Culture, Sports, Science, and
Technology of Japan and the Japan Society for the Promotion
of Science; the Australian Research Council and the
Australian Department of Education, Science and Training;
the National Natural Science Foundation of China under
contract No.~10575109 and 10775142; the Department of
Science and Technology of India; 
the BK21 program of the Ministry of Education of Korea, 
the CHEP src program and Basic Research program (grant 
No. R01-2005-000-10089-0, R01-2008-000-10477-0) of the 
Korea Science and Engineering Foundation;
the Polish State Committee for Scientific Research; 
the Ministry of Education and Science of the Russian
Federation and the Russian Federal Agency for Atomic Energy;
the Slovenian Research Agency;  the Swiss
National Science Foundation; the National Science Council
and the Ministry of Education of Taiwan; and the U.S.\
Department of Energy.


\begin{thebibliography}{99}
\bibitem{dzb}
  A small amount of data is reprocessed using a set of constants which
  is corrected for the small relative misalignment of the SVD and CDC.
  The correction function is calculated from comparisons 
  of the vertex positions for tracks between the two reprocessings.
\end{thebibliography}
\end{document}